\documentclass[fleqn,twoside]{article}
\usepackage{espcrc2}
\usepackage{amsmath,amstext,amsfonts,amsbsy,amssymb,amscd,bbm,epsfig}
\usepackage{graphicx}
\usepackage[figuresright]{rotating}

\renewcommand{\AmS}{{\protect\the\textfont2
  A\kern-.1667em\lower.5ex\hbox{M}\kern-.125emS}}

\newcommand{\ba}{\begin{array}}
\newcommand{\ea}{\end{array}}

\newcommand{\req}[1]{Eq.~(\ref{#1})}

\newcommand{\rep}[1]{\cite{#1}}
\newcommand{\refig}[1]{Fig.~\ref{#1}}

\newcommand{\dif}{{\rm d}}

\newcommand{\Dslash}{\relax{\kern+.25em / \kern-.70em D}}

\newcommand{\Real}{\relax{\mathsf{\Gamma\kern-.35em R}}}
\newcommand{\Int}{\relax{\mathsf{Z\kern-.40em Z}}}

\newcommand{\ihalf}{{\scriptstyle{{i\over 2}}}}

\newcommand{\gbar}{\kern1pt\overline{\kern-1pt g\kern-0pt}\kern1pt}
\newcommand{\mbar}{\kern2pt\overline{\kern-1pt m\kern-1pt}\kern1pt}
\newcommand{\obar}[1]{\kern3pt\overline{\kern-2pt #1\kern-0pt}\kern1pt}

\newcommand{\sigmaP}{\sigma_{\rm\scriptscriptstyle P}}
\newcommand{\sigmaT}{\sigma_{\rm\scriptscriptstyle T}}
\newcommand{\SigmaP}{\Sigma_{\rm\scriptscriptstyle P}}
\newcommand{\SigmaT}{\Sigma_{\rm\scriptscriptstyle T}}
\newcommand{\lmax}{L_{\rm max}}
\newcommand{\lmin}{L_{\rm min}}

\newcommand{\Oa}{\mbox{O}(a)}
\newcommand{\icsw}{c_{\rm sw}}
\newcommand{\ict}{c_{\rm t}}
\newcommand{\icttil}{\tilde c_{\rm t}}

\newcommand{\icT}{c_{\rm\scriptscriptstyle T}}

\newcommand{\cO}{{\cal O}}

\newcommand{\vy}{\mathbf{y}}
\newcommand{\vz}{\mathbf{z}}

\title{\vspace{-4.3cm}
       \rightline{\normalsize ROM2F/2003/21}
       \vspace{-0.1cm}
       \rightline{\normalsize MS-TP-03-11}
       \vspace{-0.1cm}
       \rightline{\normalsize DESY 03-130}
       \vspace{-0.1cm}
       \rightline{\normalsize September 2003}
       \vspace{2.0cm}
       Quark bilinear step scaling functions and their continuum limit
       extrapolation\thanks{Talk presented by C.~Pena at Lattice 2003,
       Tsukuba, Japan.}
        \setcounter{footnote}{0}
      }

\author{M.~Guagnelli\address[ToV]{INFN Sezione di Roma II,
        c/o Dipartimento di Fisica, Universit\`a di Roma ``Tor Vergata'',\\
        \ Via della Ricerca Scientifica 1, I-00133 Rome, Italy},
        J.~Heitger\address[WWUM]{Westf\"alische Wilhelms-Universit\"at
        M\"unster, Institut f\"ur Theoretische Physik,\\
        ~\,Wilhelm-Klemm-Strasse 9, D-48149 M\"unster, Germany},
        C.~Pena\address[DESY-HH]{DESY, Theory Group,
        Notkestrasse 85, D-22603 Hamburg, Germany}
        and
        A.~Vladikas\addressmark[ToV]
        (ALPHA Collaboration)
       }

\begin{document}

\begin{abstract}
Some new results on nonperturbative renormalisation of quark bilinears
in quenched QCD with Schr\"odinger Functional techniques are
presented. Special emphasis is put on a study of the universality of the
continuum limit for step scaling functions computed with different levels
of $\Oa$ improvement.
\end{abstract}

\maketitle

\section{Step scaling functions of quark bilinears}
Nonperturbative renormalisation is a key ingredient in precision
lattice QCD computations.  Schr\"odinger Functional (SF) techniques
enable us to determine nonperturbatively the renormalisation group
(RG) running of fundamental parameters (coupling and quark masses) and
composite operators in QCD over a vast range of scales, thus providing a good
control of systematics in the renormalisation procedure~\rep{Sommer:2002en}. One important
issue in this context is the extrapolation to the continuum limit (CL)
required to remove the cutoff dependence in the RG running. Whenever
this dependence is steep and cannot be tamed by appropriate $\Oa$
improvement of the action and/or the relevant operators, the
extrapolation may turn into a major source of uncertainty. Good
control of the latter is also interesting in order to test
universality of the CL for different lattice regularisations, along
the lines of e.g.~\rep{deDivitiis:1994yz}. Here we report on an
ongoing detailed investigation of the CL extrapolation in the
determination of the RG running of quark masses and quark bilinears in
quenched QCD with Wilson fermions.

To describe the nonperturbative RG running of a bilinear
$O_\Gamma=\bar\psi_i\Gamma\psi_j$, where $i,j$ are different quark flavours, we
use a step scaling function (SSF) $\sigma_\Gamma$, defined as
\begin{gather}
  \sigma_\Gamma(u) = \exp\left\{-\int_{\gbar(\mu)}^{\gbar(\mu/2)}
  \relax{\kern-10pt}\dif g \, \frac{\gamma(g)}{\beta(g)}\right\}_{u=\gbar^2(\mu)} \ ,
\end{gather}
where
$\gamma$ 
is the anomalous dimension of the operator,
$\beta(\gbar(\mu))=\partial\gbar(\mu)/\partial(\ln\mu)$, and $\gbar$ is the
renormalised coupling. In order to compute a SSF in a SF scheme, the
theory is put to live in a finite box of size $L^3 \times T$ with SF boundary
conditions\footnote{We will always take $T=L$, vanishing boundary
gauge fields and a phase $\theta=0.5$ in spatial boundary conditions.}.
The renormalisation scheme is fixed by setting $\mu=1/L$ and imposing
in the chiral limit a renormalisation condition of the
form~\rep{Capitani:1998mq}:
\begin{gather}
  \label{eq:renormalisation_condition}
  Z_\Gamma\left(g_0^2,\frac{a}{L}\right)\frac{F(T/2)}{\Theta} =
  \left.\frac{F(T/2)}{\Theta}\right|_{\rm tree~level} \ ,
\end{gather}
where $F$ is a suitable SF correlation function, and $\Theta$ is a
factor (see below for its precise form) which cancels the
multiplicative renormalisation of SF boundary fields.

The numerical determination of the SSF at a given value of $u=\gbar^2$
starts with the computation of the renormalisation constant on
lattices of sizes $L/a$ and $2L/a$ at fixed bare coupling; this allows
to construct the finite-cutoff SSF
$\Sigma_\Gamma(u,a/L)=Z_\Gamma(g_0^2,a/2L)/Z_\Gamma(g_0^2,a/L)$.
Then $\sigma_\Gamma$ is obtained by repeating the procedure at
various values of $a$ for fixed $L$,
and extrapolating to the CL:
\begin{gather}
  \sigma_\Gamma(u) = \left.\lim_{a/L \to 0}
    \Sigma_\Gamma(u,a/L)\right|_{\gbar^2(L^{-1})=u} \ .
\end{gather}

We will concentrate on two bilinear operators, namely $P \equiv O_{\gamma_5}$ and
$T_{0k} \equiv i\,O_{\sigma_{0k}}$, with
$\sigma_{0k}=\ihalf[\gamma_0,\gamma_k]$. To compute renormalisation
constants,~\req{eq:renormalisation_condition} is imposed with $F$ and
$\Theta^2$ set to
\begin{align}
  f_{\rm P}(x_0) &= -\frac{1}{2}\langle
  \bar\psi_i(x)\gamma_5\psi_j(x)\cO_{ji}[\gamma_5]
  \rangle \ , \\
  f_1 &= -\frac{1}{2L^6}\langle
  \cO'_{ij}[\gamma_5]\cO_{ji}[\gamma_5]
  \rangle
\end{align}
for the pseudoscalar density and
\begin{align}
  k_{\rm T}(x_0) &= -\frac{1}{6}\sum_k\langle
  \bar\psi_i(x)\sigma_{0k}\psi_j(x)\cO_{ji}[\gamma_k]
  \rangle \ , \\
  k_1 &= -\frac{1}{6L^6}\sum_k\langle
  \cO'_{ij}[\gamma_k]\cO_{ji}[\gamma_k]
  \rangle
\end{align}
for the tensor bilinear, where
$\cO_{ij}[\Gamma]=a^6\sum_{\vy,\vz}\bar\zeta_i(\vy)\Gamma\zeta_j(\vz)$ is a
SF boundary source. The resulting renormalisation constants and SSFs will
bear the labels P and T, respectively.

We stress that the computation of $\sigmaP$ and $\sigmaT$ suffices
to determine the RG running of all quark bilinears in the
CL. Recall also that by using axial Ward-Takahashi identities
it is easy to show that $\sigmaP=\sigma_m^{-1}$, where $\sigma_m$ is the SSF
for quark masses.

\section{Cutoff effects in step scaling functions}

Cutoff effects in $\SigmaP$ and $\SigmaT$ can be analysed by
considering the Symanzik expansion of the two-point functions entering
their definitions. In the absence of $\Oa$ improvement both quantities
are hence predicted to exhibit leading cutoff
effects linear in $a$. To implement complete on-shell $\Oa$ improvement at the
action level, it is enough to include the clover term proportional to
$\icsw$ and the boundary counterterms proportional to $\ict$ and
$\icttil$~\rep{Luscher:1996sc}. $\Oa$ improvement for the operators in the
chiral limit can be written schematically as
\begin{gather}
  \label{improved_opers}
  P^I = P~;~~~
  T_{0k}^I = T_{0k} + a\,\icT\left[\partial_0 V_k-\partial_k
  V_0\right] \ ,
\end{gather}
with $V_\mu=\bar\psi_i\gamma_\mu\psi_j$.
The coefficient $\icsw$ is known nonperturbatively in the whole range of
values of $g_0$ needed, while $\ict$ and $\icttil$ are known in
perturbation theory to $g_0^4$ and $g_0^2$ orders,
respectively~\rep{imp_coeffs}; $\icT$, on the other hand, has
been computed nonperturbatively only at a few values of
$g_0$~\rep{Bhattacharya:2000pn}.

We have computed the SSFs $\SigmaP$ and $\SigmaT$ in
quenched QCD at 14 different values of the renormalised coupling, ranging from
$\gbar^2(\lmax^{-1})=3.48$ to $\gbar^2(\lmin^{-1})=0.8873$, and four
different values of the bare coupling constant in each case (corresponding
to $L/a=6,8,12,16$). The details are essentially the same as
in~\rep{Capitani:1998mq}. The main novelty is that we have used two
different regularisations, involving different levels of $\Oa$ improvement,
the results of which will be referred to as [I] and [U], respectively:
\begin{itemize}
  \item[$\rm{[I]}$] Clover action with nonperturbative $\icsw$, one-loop values
  for $\ict$ and $\icttil$.
  \vspace{-3pt}
  \item[$\rm{[U]}$] Wilson action ($\icsw=0$), $\icttil=1$ (tree level value),
  one-loop value for $\ict$.
\end{itemize}
$\SigmaT$ has been always computed with an unimproved $T_{0k}$.

\begin{figure}[t!]
\vspace{3.7cm}
\includegraphics{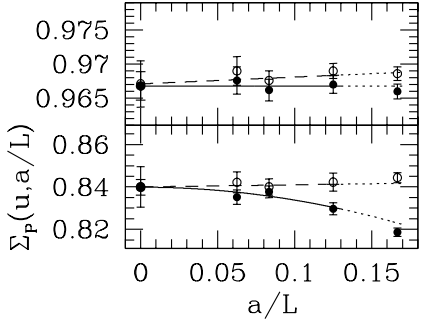}
\caption{
CL extrapolation of $\SigmaP$ for $\gbar^2=0.8873$ (top) and $\gbar^2=3.48$
(bottom). Full (open) symbols correspond to [I] ([U])
data. Fits, excluding $L/a=6$ points, have been performed with
Ans\"atze of the form $\sigma+\rho(a/L)^2$ and $\sigma+\rho(a/L)$, respectively.
}
\vspace{-7truemm}
\label{fig:sigmap_extrap}
\end{figure}

As the effect of neglecting $O(g_0^4)$ terms in $\ict$ and $\icttil$ can
be expected to result in very small $\Oa$ effects in the
SSFs~\rep{Capitani:1998mq}, the results for $\SigmaP^{\rm [I]}$ should
approach the continuum limit as $a^2$; $\SigmaP^{\rm [U]}$, on the
other hand, is expected to exhibit a linear behaviour in $a$. This
is indeed observed in numerical results, as exemplified
by~\refig{fig:sigmap_extrap}, where the CL extrapolation for $\sigmaP$
at two different values of the renormalised coupling is shown. By
fitting [I] and [U] results independently with various Ans\"atze,
universality of the CL has been checked, up to deviations which
never exceed $1.5\,\sigma$ and can be regarded as
statistical. Eventually, [I] and [U] results can be combined into a
constrained fit to obtain a more precise determination of $\sigmaP$,
as advocated in~\rep{two_actions} for computations where $\Oa$
improvement of operators is difficult to implement.

\begin{figure}[h!]
\vspace{3.7cm}
\includegraphics{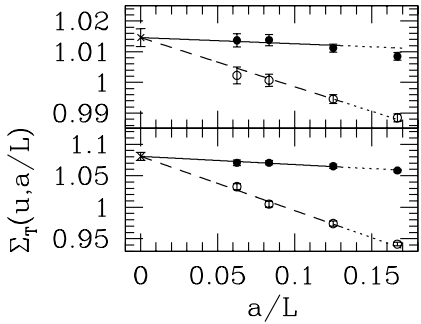}
\caption{
Same as~\refig{fig:sigmap_extrap} for $\SigmaT$, showing a constrained
CL extrapolation. Linear behaviour in $(a/L)$ has been assumed
both for [I] and [U] data.
}
\vspace{-5truemm}
\label{fig:sigmat_extrap}
\end{figure}

\begin{figure}[t!]
\vspace{3.9cm}
\includegraphics{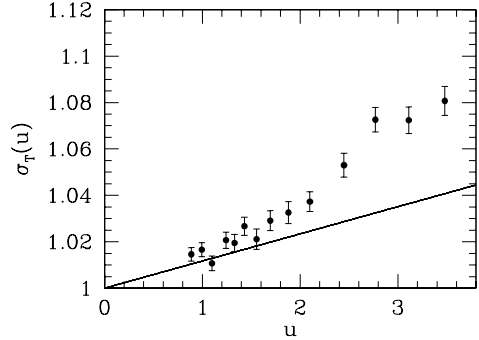}
\caption{
Preliminary results for $\sigmaT$ as a function of $u=\gbar^2$ from a
constrained fit. The solid line is the one-loop result for $\sigmaT$.
}
\vspace{-6truemm}
\label{fig:sigmat}
\end{figure}

$\SigmaT$ is expected, on the other hand, to exhibit linear
cutoff effects even when it is computed with an $\Oa$ improved action, due
to the lack of operator improvement. Furthermore, $\SigmaT^{\rm [U]}$ may
show large $\Oa$ effects. This is indeed observed in preliminary
results for this SSF (\refig{fig:sigmat_extrap}). In this case, the
two-action strategy reveals its usefulness, as assuming universality of the
CL and performing a constrained fit yields much more precise CL values than
would follow from [I] data alone. The corresponding result
for $\sigmaT$ is depicted in~\refig{fig:sigmat}.

\section{Outlook}

The results obtained so far demonstrate universality of the
CL result for the RG running of quark bilinears computed with
actions with different levels of $\Oa$ improvement. A better control
of the CL extrapolation can be achieved in cases where
nonperturbatively $\Oa$ improved operators are not available
by combining results from different actions. This
technique has been used in our preliminary computation of the
nonperturbative RG running of tensor bilinears in quenched
QCD. Pending tasks include comparison with
cutoff effects in perturbation theory, and the computation of the NLO
anomalous dimension of $T_{0k}$, required to match our results for
$\sigmaT$ to conventional renormalisation schemes.

\section*{Acknowledgements}
\noindent Work supported in part by the European Union's Human
Potential Programme under contract HPRN-CT-2000-00145, Hadrons/Lattice
QCD.


\begin{thebibliography}{9}

\bibitem{Sommer:2002en}
For a recent review see R.~Sommer,
Nucl. Phys. Proc. Suppl. 119 (2003) 185.

\bibitem{deDivitiis:1994yz}
ALPHA, G.~de Divitiis {\it et al.}, 
Nucl. Phys. B437 (1995) 447.

\bibitem{Capitani:1998mq}
ALPHA,
S.~Capitani et al., 
Nucl. Phys. B544 (1999) 669.

\bibitem{Luscher:1996sc}
M.~L\"uscher et al.,
Nucl. Phys. B478 (1996) 365.

\bibitem{imp_coeffs}
M.~L\"uscher et al.,
Nucl. Phys. B491 (1997) 323;
ALPHA, A.~Bode, U.~Wolff and P.~Weisz, 
Nucl. Phys. B540 (1999) 491;
M.~L\"uscher and P.~Weisz,
Nucl. Phys. B479 (1996) 429.

\bibitem{Bhattacharya:2000pn}
T.~Bhattacharya et al.,
Phys. Rev. D63 (2001) 074505.

\bibitem{two_actions}
M.~Guagnelli, K.~Jansen and R.~Petronzio,
Phys. Lett. B457 (1999) 153;
ALPHA,
M.~Guagnelli et al., 
Nucl. Phys. Proc. Suppl. 119 (2003) 436.

\end{thebibliography}
\end{document}